\documentclass[11pt]{article}  	
\usepackage{geometry}                		
\usepackage{graphicx}				

\usepackage{amssymb}
\usepackage{amsmath}
\usepackage[utf8]{inputenc}
\usepackage{multirow}
\usepackage{array}
\usepackage{mathtools}
\usepackage{makecell}
\usepackage{diagbox}
\usepackage{soul}
\usepackage[all,knot,poly]{xy}
\usepackage{color}
\numberwithin{equation}{section}
\usepackage{multicol,caption}
\usepackage{empheq}
\usepackage{enumitem}
\usepackage{verbatim}
\usepackage{microtype}
\usepackage[retainorgcmds]{IEEEtrantools}

\title{On the SLq(2) Extension of the Standard Model and the Measure of Charge}
\author{Robert J. Finkelstein}
\date{\emph{Physics \& Astronomy, University of California, Los Angeles, \\
475 Portola Plaza, Los Angeles, California 90095, USA \\
finkel@physics.ucla.edu}}

\begin{document}

\maketitle

\begin{abstract}
Our SLq(2) extension of the standard model is constructed by replacing the elementary field operators, $\Psi (x)$, of the standard model by $\hat{\Psi}^{j}_{mm'}(x) D^{j}_{mm'}$ where $D^{j}_{mm'}$ is an element of the $2j + 1$ dimensional representation of the SLq(2) algebra, which is also the knot algebra. The allowed quantum states $(j,m,m')$ are restricted by the topological conditions
\begin{equation*}
(j,m,m') = \frac{1}{2}(N,w,r+o)
\end{equation*}
postulated between the states of the quantum knot $(j,m,m')$ and the corresponding classical knot $(N,w,r+o)$ where the $(N,w,r)$ are (the number of crossings, the writhe, the rotation) of the 2d projection of the corresponding oriented classical knot. Here $o$ is an odd number that is required by the difference in parity between $w$ and $r$. There is also the empirical restriction on the allowed states
\begin{equation*}
(j,m,m')=3(t,-t_3,-t_0)_L
\end{equation*}
that holds at the $j=\frac{3}{2}$ level, connecting quantum trefoils $(\frac{3}{2},m,m')$ with leptons and quarks $(\frac{1}{2}, -t_3, -t_0)_L$. The so constructed knotted leptons and quarks turn out to be composed of three $j=\frac{1}{2}$ particles which unexpectedly agree with the preon models of Harrari and Shupe. The $j=0$ particles, being electroweak neutral, are dark and plausibly greatly outnumber the quarks and leptons. The SLq(2) or $(j,m,m')$ measure of charge has a direct physical interpretation since $2j$ is the total number of preonic charges while $2m$ and $2m'$ are the numbers of writhe and rotation sources of preonic charge. The total SLq(2) charge of a particle, measured by writhe and rotation and composed of preons, sums the signs of the counterclockwise turns $(+1)$ and clockwise turns $(-1)$ that any energy-momentum current makes in going once around the knot. In this way the handedness of the knot reduces charge to a geometric concept similar to the way that curvature of spacetime encodes mass and energy. According to this model, the leptons and quarks are $j=\frac{3}{2}$ particles, the preons are $j=\frac{1}{2}$ particles, and the $j=0$ particles are candidates for dark matter. It is possible to understand $q$ as a simple deformation parameter or as the ratio $e/g$ of the electroweak to the gluon coupling constants.
\\
\emph{Keywords:} Quantum groups; electroweak; knot models; preon models; dark matter. \\
PACS numbers: 02.20.Uw, 02.10.Kn, 12.60.Fr
\end{abstract}

\section{Introduction}

In our SLq(2) extension of the standard model or similar models the field operators $\Psi (x)$ are replaced in the following way:$^{1-3}$
\begin{equation}
\Psi (x) \rightarrow \hat{\Psi}^{j}_{mm'}(x) D^{j}_{mm'}
\end{equation}
where $D^{j}_{mm'}$ is an element of the $2j+1$ dimensional representation of the SLq(2) algebra and where the $\hat{\Psi}^{j}_{mm'}(x)$ satisfy the standard Lagrangians after modification by the form factors generated by the $D^{j}_{mm'}$. Since SLq(2) describes the symmetry of the classical knot, the field quanta, lying in the SLq(2) algebra, may be characterized as knotted or may be described as quantum knots, similar to the way that spin is introduced by the $D_{mm'}^j$ of the rotation group.

We further define the quantum knot by \emph{postulating} the following kinematical restriction to \emph{allowed quantum states}
\begin{equation}
(j,m,m') = \frac{1}{2}(N, w, r+o)
\end{equation}
where $N$ is the number of crossings, $w$ is the writhe, $r$ is the rotation of the \emph{2d projection} of the corresponding oriented classical knot; and $o$ is an odd integer that is required since $w$ and $r$ are of opposite parity, while $2m$ and $2m'$ are of the same parity. This equation establishes a correspondence between \emph{states of the quantum knot and the topology of a classical knot.} 

The writhe counts the turns of the tangent at the crossings and the rotation counts the turns of the tangent in one complete circuit of the knot. Here $o$ is a new quantum number which we set at $o=1$ for the simplest knot, the quantum trefoil, and may have other values for other knots. We also introduce ``the quantum rotation" $\tilde{r} \equiv r+o$. Then $o$ appears as a zero-point rotation of the quantum knot. We now write
\begin{equation}
D_{mm'}^j = D_{\frac{w}{2} \frac{r+o}{2}}^{N/2} \tag*{$(1.2)'$}
\end{equation}
for quantum knots where $(N,w,r)$ describe the topology of the corresponding classical knot and highly restrict the quantum kinematics.
 
The knot picture of the elementary particles is more attractive if the simplest particles are the simplest knots. We therefore consider the possibility that the most elementary fermions with electroweak isotopic spin $t=\frac{1}{2}$ are the most elementary quantum knots, namely the quantum trefoils with $N=3$ and $o=1$. This possibility is supported by the following empirical observation
\begin{equation}
(t, -t_3, -t_0)_L = \frac{1}{6}(N, w, r+1) \label{fourleftchiral}
\end{equation}
where $t$ and $t_3$ refer to isotopic spin and $t_0$ is the electroweak hypercharge. Equation \eqref{fourleftchiral} relates the four left chiral families of the \emph{elementary fermions} described by $(\frac{1}{2}, t_3, t_0)_L$ to the \emph{2d-projections of the four quantum trefoils} described by $(3, w, r+1)$ and is supported by the row to row proportionality in Table 1 as expressed in equation \eqref{fourleftchiral}.

\renewcommand{\arraystretch}{2}
\begin{center}
\begin{tabular}{r c r r r | l c r r c}
\multicolumn{10}{c}{\textbf{Table 1:} Empirical Support for \eqref{fourleftchiral}} \\
\hline \hline
 & $(f_1, f_2, f_3)$ & $t$ & $t_3$ & $t_0$ & $D^{N/2}_{\frac{w}{2}\frac{r+1}{2}}$ & $N$ & $w$ & $r$ & $r+1$ \\[0.2cm]
\hline
\multirow{2}{*}{\hspace{-4pt}leptons $\Bigg \{$} & $(e, \mu, \tau)_L$ & $\frac{1}{2}$ & $-\frac{1}{2}$ & $-\frac{1}{2}$ & $D^{3/2}_{\frac{3}{2}\frac{3}{2}}$ & 3 & 3 & 2 & 3 \\
& $(\nu_e, \nu_{\mu}, \nu_{\tau})_L$ & $\frac{1}{2}$ & $\frac{1}{2}$ & $-\frac{1}{2}$ & $D^{3/2}_{-\frac{3}{2}\frac{3}{2}}$ & 3 & $-3$ & 2 & 3 \\
\multirow{2}{*}{quarks $\Bigg \{$} & $(d, s, b)_L$ & $\frac{1}{2}$ & $-\frac{1}{2}$ & $\frac{1}{6}$ & $D^{3/2}_{\frac{3}{2}-\frac{1}{2}}$ & 3 & 3 & $-2$ & $-1$ \\
& $(u, c, t)_L$ & $\frac{1}{2}$ & $\frac{1}{2}$ & $\frac{1}{6}$ & $D^{3/2}_{-\frac{3}{2}-\frac{1}{2}}$ & 3 & $-3$ & $-2$ & $-1$ \\ [0.2cm]
\hline
\end{tabular}
\end{center}

Only for the particular row-to-row correspondences shown in Table 1 does \eqref{fourleftchiral} hold, i.e., \emph{each of the four families of fermions labelled by $(t_3, t_0)$ is uniquely correlated with a specific $(w,r)$ classical trefoil, and therefore with a specific state $D^{N/2}_{\frac{w}{2}\frac{r+1}{2}}$ of the quantum trefoil.}

Note that the $t_3$ doublets now become the writhe doublets ($w = \pm 3$). Note also that with this same correspondence the leptons and quarks form a knot rotation doublet ($r = \pm 2$).

One now has for $j = \frac{3}{2}$, $t = \frac{1}{2}$ and $N=3$
\begin{IEEEeqnarray*}{rClsr}
(j,m,m') & = & \frac{1}{2}(N, w, r+o) & \quad (postulated relation between quantum and classical & \\
\IEEEeqnarraymulticol{4}{u}{knots)} & \quad (1.2) \\
(t, -t_3, -t_0)_L &= & \frac{1}{6}(N, w, r+1) & \quad (empirical relation between elementary fermions & \\
\IEEEeqnarraymulticol{4}{u}{and classical trefoils)} & \quad (1.3) \\
\IEEEeqnarraymulticol{5}{s}{Then by (1.2) and (1.3) one also has} \\
(j,m,m') & = & 3(t, -t_3, -t_0)_L & \quad (empirically based relation between quantum trefoils & \\ \IEEEeqnarraymulticol{4}{u}{and elementary fermions)} & \quad (1.4)
\end{IEEEeqnarray*}
for the left chiral fields and quantum trefoils. 

It is now proposed that the elementary fermions are $(j, m, m')$ states of SLq(2) satisfying (1.2) and (1.4).

With the aid of Noether charges carried by $D^j_{mm'}$ knots we shall next explore the extension of (1.2) and (1.4) to the more general case where $t \neq 1/2$, $j \neq 3/2$ and $o \neq 1$.

\section{Noether Charges carried by $D^j_{mm'}$ Knots} 
The $2j+1$ dimensional representations of SLq(2) may be expressed as $\text{follows}^{(3)}$
\begin{align}
D^j_{mm'} (q \vert a, b, c, d) = \sum_{\substack{\delta(n_a + n_b, n_+) \\ \delta(n_c + n_d, n_-)}} A^j_{mm'} (q \vert n_a, n_c) \delta(n_a + n_c, n_+') a^{n_a} b^{n_b} c^{n_c} d^{n_d}
\end{align}
The sum (2.1) is taken over the positive integers $n_a, n_b, n_c, n_d$ subject to the $\delta$-function constraints as shown. $q$ is taken to be real. Here the arguments $(a,b,c,d)$ satisfy the following algebra 
\begin{equation}
\begin{split}
ab = qba \qquad bd = qdb \qquad ad-qbc = 1 \qquad bc &= cb \\
ac = qca \qquad cd = qdc \qquad da -q_1cb = 1 \qquad q_1 &\equiv q^{-1}
\end{split}
\end{equation}
The numerical coefficients are
\begin{align}
A^j_{mm'}(q \vert n_a, n_c) = \bigg[\frac{\langle n_+' \rangle_1 \langle n_-' \rangle_1}{\langle n_+ \rangle_1 \langle n_- \rangle_1} \bigg]^{\frac{1}{2}} \frac{\langle n_+ \rangle_1!}{\langle n_a \rangle_1! \langle n_b \rangle_1!} \frac{\langle n_- \rangle_1!}{\langle n_c \rangle_1! \langle n_d \rangle_1!}
\end{align}
where
\begin{align}
\begin{tabular}{c}
$n_{\pm} = j \pm m$ \\
$n_{\pm}' = j \pm m'$
\end{tabular}
\end{align}
and
\begin{align}
\begin{tabular}{c}
$\langle n \rangle_q = \frac{q^n -1}{q-1}$ \\
$\langle n \rangle_1 \equiv \langle n \rangle_{q_1}$
\end{tabular}
\end{align}
The algebra (2.2) is invariant under the following gauge transformation:
\begin{align}
\begin{tabular}{l l}
$a' = e^{i\varphi_a}a$ & $b' = e^{i\varphi_b}b$ \\
$d' = e^{-i\varphi_a}d$ & $c' = e^{-i\varphi_b}c$
\end{tabular}
\end{align}
We shall also refer to the transformation described by (2.6) as $U_a(1) \times U_b(1)$.

The transformation, $U_a(1) \times U_b(1)$, on the $(a,b,c,d)$ of SLq(2) induces on the $D^j_{mm'}$ of SLq(2) the corresponding transformation$^{(3)}$ 
\begin{align*}
\begin{tabular}{r l}
$D^j_{mm'} (a,b,c,d) \rightarrow$ & $D^j_{mm'} (a',b',c',d')$ \\
= & $e^{i(\varphi_a + \varphi_b)m}e^{i(\varphi_a - \varphi_b)m'}D^j_{mm'}(a,b,c,d)$ \\
= & $U_m(1) \times U_{m'}(1)D^j_{mm'}(a,b,c,d)$ \tag*{$(2.6)'$} 
\end{tabular}
\end{align*}
and on the field operators modified by the $D^j_{mm'}$
\begin{align}
\Psi^j_{mm'} \rightarrow U_m(1) \times U_{m'}(1) \Psi^j_{mm'}
\end{align}

\emph{For physical consistency any allowed field action must be invariant under (2.7) since (2.7) is induced by $U_a \times U_b$ transformations that leave the defining algebra (2.2) unchanged} \footnote{The field action of the extended standard model is invariant under (2.7) because of the hermitian structure of the standard model, and adjustment of the Fermion-Boson interaction.}. 
\emph{There are then Noether charges associated with $U_m$ and $U_{m'}$ that may be described as writhe and rotation charges, $Q_w$ and $Q_r$, since $m=\frac{w}{2}$ and $m'=\frac{1}{2}(r+o)$ for quantum knots.}

For quantum trefoils we have set $o=1$, and we now define their Noether charges:
\begin{align}
Q_w \equiv -k_wm \quad \left(\equiv -k_w\frac{w}{2} \right)
\end{align}
\begin{align}
Q_r \equiv -k_rm' \quad \left(\equiv -k_r\frac{1}{2}(r+1) \right)
\end{align}
where $k_w$ and $k_r$ are undetermined constants. 

Retaining the row-row correspondence established in Table 1, we next compare in Table 2 the electroweak charges $Q_e$ of the elementary fermions with the total Noether charges of the corresponding quantum trefoils.

\begin{center}
\begin{tabular} {c r r r c | c l c c c}
\multicolumn{10}{c}{{\textbf{Table 2:} Electric Charges on Leptons, Quarks, and Quantum Trefoils}} \\
\hline \hline
\multicolumn{5}{c |}{{Standard Model}} & \multicolumn{5}{c}{{Quantum Trefoil Model}} \\
\hline
{$(f_1, f_2, f_3)$} & {$t$} & {$t_3$} &{$t_0$} & {$Q_e$} & {$(N,w,r)$} & {$D^{N/2}_{\frac{w}{2}\frac{r+1}{2}}$} & {$Q_w$} & {$Q_r$} & {$Q_w + Q_r$} \\ [0.2cm]
\hline
$(e, \mu, \tau)_L$ & $\frac{1}{2}$ & $-\frac{1}{2}$ & $-\frac{1}{2}$ & $-e$ & $(3,3,2)$ & $D^{3/2}_{\frac{3}{2} \frac{3}{2}}$ & $-k_w \left( \frac{3}{2} \right)$ & $-k_r \left( \frac{3}{2} \right)$ & $-\frac{3}{2}(k_r + k_w)$ \\
$(\nu_e, \nu_{\mu}, \nu_{\tau})_L \hspace{-10pt}$ & $\frac{1}{2}$ & $\frac{1}{2}$ & $-\frac{1}{2}$ & $0$ & $(3,-3,2)$ & $D^{3/2}_{-\frac{3}{2} \frac{3}{2}}$ & $-k_w \left( -\frac{3}{2} \right)$ & $-k_r \left( \frac{3}{2} \right)$ & $\frac{3}{2}(k_w - k_r) $ \\
$(d,s,b)_L$ & $\frac{1}{2}$ & $-\frac{1}{2}$ & $\frac{1}{6}$ & $-\frac{1}{3}e$ & $(3,3,-2)$ & $D^{3/2}_{\frac{3}{2} -\frac{1}{2}}$ & $-k_w \left( \frac{3}{2} \right)$ & $-k_r \left( -\frac{1}{2} \right)$ & $\frac{1}{2}(k_r - 3k_w)$ \\
$(u,c,t)_L$ & $\frac{1}{2}$ & $\frac{1}{2}$ & $\frac{1}{6}$ & $\frac{2}{3}e$ & $(3,-3, -2)$ & $D^{3/2}_{-\frac{3}{2} -\frac{1}{2}} \hspace{-5pt}$ & $-k_w \left( -\frac{3}{2} \right)$ & $-k_r \left( -\frac{1}{2} \right)$ & $\frac{1}{2}(k_r + 3k_w)$ \\
& & \multicolumn{3}{c |}{$ \hspace{-8pt} Q_e = e(t_3+t_0) \hspace{-10pt}$} & \multicolumn{2}{c}{\normalsize $(j, m, m') = \frac{1}{2}(N, w, r + 1)$} & $Q_w = -k_w \frac{w}{2} \hspace{-5pt}$ & $Q_r = -k_r \frac{r+1}{2} \hspace{-15pt}$ & \\
\hline
\end{tabular}
\end{center}

One sees that $Q_w + Q_r = Q_e$ for charged leptons, neutrinos and for both up and down quarks if 
\begin{align}
k_r = k_w = k=\frac{e}{3}
\end{align}
and also that $t_3$ and $t_0$ measure the writhe and rotation charges respectively:
\begin{align}
Q_w = et_3
\end{align}
\begin{align}
Q_r = e t_0
\end{align}
Then $Q_w + Q_r = Q_e$ becomes by (2.11) and (2.12) an alternative statement of
\begin{align}
Q_e = e(t_3 + t_0)
\end{align}
of the standard model.

In SLq(2) measure $Q_e = Q_w + Q_r$ is by $(2.8)$ and $(2.9)$: 
\begin{align}
Q_e = -\frac{e}{3}(m+m'),
\end{align}
or
\begin{align}
Q_e = - \frac{e}{6}(w+r+1).
\end{align}
for the quantum trefoils, that represent the elementary fermions.

Then the electric charge is a measure of the writhe $+$ rotation of the trefoil. The total electric charge in this way resembles the total angular momentum as a sum of two parts where the knot rotation corresponds to the orbital angular momentum and where the localized contribution of the writhe to the charge corresponds to the localized contribution of the spin to the angular momentum. In (2.15) $o=1$ contributes a ``zero-point charge."

The total SLq(2) charge sums the signed clockwise and counterclockwise turns that any energy-momentum current makes both at the crossings and in going once around the 2d-projected knot. In this way, the ``handedness" of the knot determines its charge, so that handedness reduces charge to a geometrical concept similar to the way that curvature of space-time geometrizes mass and energy. This measure of charge, which is suggested by the leptons and quarks, goes to a deeper level than the electroweak isotopic measure that originated in the approximate equality of masses in the neutron-proton system.

As here defined, quantum knots carry the charge expressed as both $t_3 + t_0$ and $m + m'$. The conventional \textbf{$\bf{(t_3, t_0)}$ measure of charge is based on $\bf{SU(2) \times U(1)}$ while the $\bf{(m,m')}$ measure of charge is based on SLq(2)}. These two different measures are related at the $j=\frac{3}{2}$ level by $(j,m,m') = 3(t,-t_3,-t_0)$. We shall next attempt to extend this analysis beyond $j=\frac{3}{2}$, and in particular to the fundamental representation $j=\frac{1}{2}$.

\section{The Preon Interpretation of $D^j_{mm'}$}
We may try to give physical meaning to the defining expression for $D_{mm'}^j$,
\begin{align*}
D^j_{mm'} (q \vert a, b, c, d) = \sum_{\substack{\delta(n_a + n_b, n_+) \\ \delta(n_c + n_d, n_-)}} A^j_{mm'} (q \vert n_a, n_c) \delta(n_a + n_c, n_+') a^{n_a} b^{n_b} c^{n_c} d^{n_d} \tag{2.1}
\end{align*}
\emph {by interpreting the $a,b,c,d$ as creation operators for fermionic preons}, since these are also the four elements of the \emph{fundamental} $(j = \frac{1}{2})$ \emph{representation} given by:
\begin{align}
D^{1/2}_{mm'} = 
\begin{tabular}{r | l r}
\theadfont\diagbox{m}{m'} & $\frac{1}{2}$ & $- \frac{1}{2}$ \\
\hline
$\frac{1}{2}$ & $a$ & $b$ \\
$-\frac{1}{2}$ & $c$ & $d$ \\
\end{tabular}
\end{align}
By extension of (2.14) from $j = \frac{3}{2}$ to $j = \frac{1}{2}$ the Noether charge is
\begin{align}
Q_e = -\frac{e}{3}(m+m') \tag{2.14}
\end{align}

\emph{By (3.1) and (2.14) there is one charged preon, a, with charge $-\frac{e}{3}$ and its antiparticle, d, and there is one neutral preon, b, with its antiparticle, c.}

If $j = \frac{1}{2}$, then $N=1$ by the postulate (1.2) relating to the corresponding classical knot
\begin{align*} 
(j,m,m') = \frac{1}{2}(N, w, r+o)  \tag{1.2}
\end{align*}
and the corresponding $a,b,c,d$ classical realizations of the preons cannot be described as knots since they have only a single crossing. They can, however, be described as \emph{2d-projections of twisted loops with $N=1$, $w= \pm 1$ and $r=0$}. 


Having tentatively interpreted the fundamental representation in terms of preons, we next consider the general representation.

\section*{Interpretation of all $D_{mm'}^j (q \vert a,b,c,d)$}
\emph {Every $D^j_{mm'}$, as given in}
\begin{align*}
D^j_{mm'} (q \vert a, b, c, d) = \sum_{\substack{\delta(n_a + n_b, n_+) \\ \delta(n_c + n_d, n_-)}} A^j_{mm'} (q \vert n_a, n_c) \delta(n_a + n_c, n_+') a^{n_a} b^{n_b} c^{n_c} d^{n_d} \tag{2.1}
\end{align*}
\emph{being a polynomial in $a,b,c,d$, can be interpreted as a creation operator for a superposition of states, each state with $n_a, n_b, n_c, n_d$ preons.} 

It then turns out that the creation operators \textbf{(} for the charged leptons, $D^{3/2}_{\frac{3}{2} \frac{3}{2}}$; neutrinos, $D^{3/2}_{-\frac{3}{2} \frac{3}{2}}$; down quarks, $D^{3/2}_{\frac{3}{2} -\frac{1}{2}}$; and up quarks, $D^{3/2}_{-\frac{3}{2} -\frac{1}{2}}$ \textbf{)}, \emph{as empirically required by Tables 1 and 2}, are represented by (2.1) as the following monomials

\begin{center}
\begin{tabular}{c c c c c}
$D^{3/2}_{\frac{3}{2} \frac{3}{2}} \sim a^3$ & $D^{3/2}_{-\frac{3}{2} \frac{3}{2}} \sim c^3$ & $D^{3/2}_{\frac{3}{2} -\frac{1}{2}} \sim ab^2$ & $D^{3/2}_{-\frac{3}{2} -\frac{1}{2}} \sim cd^2$ & \hspace{22pt} (3.2) \\
\text{charged leptons} & \text{neutrinos} & \text{down quarks} & \text{up quarks}
\end{tabular}
\end{center}
implying that \emph{charged leptons and neutrinos are composed of three $a$-preons and three $c$-preons, respectively, while the down quarks are composed of one $a$- and two $b$-preons, and the up quarks are composed of one $c$- and two $d$-preons in agreement with the Harari-Shupe model of quarks, and with the experimental evidence on which their model is constructed.}$^{(4)(5)}$ Note that the number of preons equals the number of crossings ($(j = \frac{N}{2} = \frac{3}{2})$ in (3.2)).
\newline

Note also that there are only four ``elementary fermions" differing by the two possibilities for the writhe and the two possibilities for the rotation. Each of the ``elementary fermions" has 3 states of excitation, determined by eigenstates of $\bar{D}^j_{mm'} D^j_{mm'}$$^{(1)}$.

\section{The Knotted Electroweak Vectors} 

To achieve the required $U_a(1) \times U_b(1)$ invariance of the knotted Lagrangian (and the associated conservation of $t_3$ and $t_0$, or equivalently of the writhe and rotation charge), it is necessary to impose topological and empirical restrictions on the knotted vector bosons by which the knotted fermions interact as well as on the knotted fermions. \emph{For these electroweak vector fields we assume the $t=1$ of the standard model and therefore $j=3$ and $N=6$, in accord with (1.4) and (1.2)} \vspace{-0.2em} 
\begin{align*}
(j,m,m') = 3(t, -t_3, -t_0) \tag{1.4} \\
(j,m,m') = \frac{1}{2}(N,w,r+o) \tag{1.2}
\end{align*}
 that hold for the elementary fermion fields and \emph{that we now assume for the knotted vector fields as shown in Table 3.}
\begin{center} \vspace{-1.1em}
\begin{tabular}{r | r r r r l}
\multicolumn{6}{c}{\textbf{Table 3:} Electroweak Vectors $(j=3)$} \\
\hline \hline
 & $Q$ & $t$ & $t_3$ & $t_0$ & $D^{3t}_{-3t_3 - 3t_0}$ \\
 \hline
 $W^+$ & $e$ & 1 & $1$ & 0 & $D^3_{-3,0} \sim c^3d^3$ \\
 $W^-$ & $-e$ & 1 & $-1$ & 0 & $D^3_{3,0} \sim a^3b^3$ \\
 $W^3$ & 0 & 1 & 0 & 0 & $D^3_{0,0} \sim f_3(bc)$ \\
 \hline
\end{tabular}
\end{center}
The charged $W^+_{\mu}$ and $W^-_{\mu}$ are six preon monomials. The neutral vector $W^3_{\mu}$ is the superposition of four states of six preons given by \vspace{-0.2em}
\begin{align}
D^3_{00} = A(0,3)b^3c^3 + A(1,2)ab^2c^2d + A(2,1)a^2bcd^2 + A(3,0)a^3d^3 \tag{4.1} \vspace{-0.6em}
\end{align}
according to (2.1) which is reducible by the algebra (2.2) to a function of the neutral operator $bc$. \vspace{-1.0em}
\newline

Table 3 again illustrates the fact that the number of crossings equals the number of preons.

\newpage

\section{The SU(3) Couplings of the Standard Model}

The previous considerations are based on electroweak physics. To describe the strong interactions it is necessary according to the standard model to introduce SU(3). In the SLq(2) electroweak model, as here described, the need for the additional SU(3) symmetry is built into the knot model already at the level of the charged leptons and neutrinos since they are presented as $a^3$ and $c^3$, respectively. \emph{Then the simple way to protect the Pauli principle is to replace} $(a,c)$ \emph{by} $(a_i, c_i)$ \emph{and antisymmetrize}


\begin{center}
\begin{tabular}{r l r}
\emph{the creation operators for charged leptons} & $a^3 \text{ by } \varepsilon^{ijk}a_i a_j a_k$ & (5.1) \\
\emph{the creation operators for neutrinos} & $c^3 \text{ by } \varepsilon^{ijk}c_i c_j c_k$ & (5.2) \\
\end{tabular}
\end{center}
\emph{where $a_i$ and $c_i$ provide a basis for the fundamental representation of SU(3). Then the charged leptons and neutrinos are color singlets. If the $b$ and $d$ preons are also color singlets, then down quarks $a_ib^2$ and up quarks $c_id^2$ provide a basis for the fundamental representation of SU(3), as required by the standard model.}$^{(6)}$

We do not depart from SLq(2) in the above way of introducing SU(3). If one instead goes over to SUq(2) where $\bar{a}=d$ and $\bar{c}=-q_1b$, we may make use of the two complex representations $3$ and $\overline{3}$ of SU(3), by assigning $a_i$ and $c_i$ to the $\overline{3}$ and $b^i$ and $d^i$ to the 3 representation.$^{(7)}$
\vspace{-.5 cm}

\section{Graphical Representation of Corresponding Classical Structures}
The representation of the four classical trefoils as composed of three overlapping preon loops is shown in Figure 1. In interpreting Figure 1, note that the two lobes of all the preon loops make opposite contributions to the rotation, $r$, so that the total rotation of each preon loop vanishes. When the three $a$-preons and $c$-preons are combined to form charged leptons and neutrinos, respectively, each of the three labelled circuits is counterclockwise and contributes $+1$ to the rotation while the single unlabeled and shared (overlapping) circuit is clockwise and contributes $-1$ to the rotation so that the total $r$ for both charged leptons and neutrinos is $+2$. \vspace{-0.3em}

For quarks the three labelled loops contribute $-1$ and the shared loop $+1$ so that $r=-2$.

In each case the three preons that form a lepton trefoil contribute their three negative rotation charges. The geometric and charge profile of the lepton trefoil is similar to the geometric and charge profile of a triatomic molecule composed of neutral atoms since the electronic charges, which cancel the nuclear charges, are shared among the atoms forming the molecule just as the negative rotation charges which cancel the positive rotation charges of the preons are shared among the preons forming the trefoils. There is a similar correspondence between quarks and antimolecules.

\begin{figure}[p] 
\begin{center}
\normalsize
\begin{tabular}{c | c} 
         \multicolumn{2}{c}{\textbf{Graphical Representation of Corresponding Classical Structures}} \\ [-0.54cm]
	\multicolumn{2}{c}{\textbf{Figure 1:} Preonic Structure of Elementary Fermions} \\ [-0.49cm]
	\multicolumn{2}{c}{$Q = -\frac{e}{6}(w+r+o)$, and $(j, m, m') = \frac{1}{2} (N, w, r+o)$} \\ [-0.25cm]
	\begin{tabular}{c c c}
		\multicolumn{3}{r}{ \ul{$(w, r, o)$}} \\ [-0.3cm]
		\multicolumn{3}{l}{Charged Leptons, $D^{3/2}_{\frac{3}{2} \frac{3}{2}} \sim a^3$} \\ [0.2cm]
		 & $\hspace{17pt} a_j$ & \\ [0.4cm]
		 $\hspace{28pt} a_i$ & & $\hspace{-58pt} a_k$ \\ [-2.8cm]
		 \multicolumn{3}{l}{\xygraph{
!{0;/r1.3pc/:}
!{ \xoverh }
[u(1)] [l(0.5)]
!{\color{black} \xbendu }
[l(2)]
!{\vcap[2] =>}
!{\xbendd- \color{black}}
[d(1)] [r(0.5)]
!{\xunderh }
[d(0.5)]
!{ \xbendr}
[u(2)]
!{\hcap[2]=>}
[l(1)]
!{\xbendl- }
[l(1.5)] [d(0.25)]
!{\xbendl[0.5]}
[u(1)] [l(0.5)]
!{\xbendr[0.5] =<}
[u(1.75)] [l(2)]
!{ \xoverv}
[u(1.5)] [l(1)]
!{\color{black} \xbendr-}
[u(1)] [l(1)]
!{ \hcap[-2]=>}
[d(1)]
!{ \xbendl \color{black}}}} \\ [-1.6cm]
\multicolumn{3}{l}{ \hspace{33pt} \textcolor{red}{
\xygraph{
!{0;/r1.3pc/:}
[u(2.25)] [r(3.25)]
!{\xcapv[1] =>}
[l(2.75)] [u(.75)]
!{\xbendr[-1] =>}
[d(1)] [r(1.25)]
!{\xcaph[-1] =>}
}}}
 \\ [-0.7cm]
		 \multicolumn{3}{r}{\hspace{150pt}$(3,2,1)$}
		 
	\end{tabular}
	&
	\begin{tabular}{c c c}
		\multicolumn{3}{r}{\ul{$(w, r, o)$}} \\ [-0.3cm]
		\multicolumn{3}{l}{$a$-preons, $D^{1/2}_{\frac{1}{2} \frac{1}{2}}$} \\ [1.58cm]
		  & &\hspace{-63pt} $a$ \\ [-1.3cm]
		 \multicolumn{3}{l}{\xygraph{
!{0;/r1.3pc/:}
!{\xoverv=>}
[u(0.5)] [l(1)]
!{\xbendl}
[u(2)]
!{\hcap[-2]}
!{\xbendr-}
[r(1)]
!{\xbendr}
[u(2)]
!{\hcap[2]}
[l(1)]
!{\xbendl-}}} \\ [-1.2cm]
\multicolumn{3}{l}{\hspace{39pt} \textcolor{red}{\xygraph{
!{0;/r1.3pc/:}
[d(0.75)] [l(0.75)]
!{\xcaph[-1]=>}
}
}} \\ [-0.7cm]
		 \multicolumn{3}{r}{\hspace{150pt}$(1,0,1)$}
		 
	\end{tabular} \\ [-0.1cm] \hline
\begin{tabular}{c c c}
		\multicolumn{3}{l}{Neutrinos, $D^{3/2}_{-\frac{3}{2} \frac{3}{2}} \sim c^3$} \\ [0.2cm]
		 & $\hspace{17pt} c_j$ & \\ [0.4cm]
		 $\hspace{28pt} c_i$ & & $\hspace{-58pt} c_k$ \\ [-2.8cm]
		 \multicolumn{3}{l}{\xygraph{
!{0;/r1.3pc/:}
!{\xunderh }
[u(1)] [l(0.5)]
!{\color{black} \xbendu}
[l(2)]
!{ \vcap[2]=>}
!{\xbendd- \color{black}}
[d(1)] [r(0.5)]
!{\xoverh}
[d(0.5)]
!{\xbendr}
[u(2)]
!{\hcap[2]=>}
[l(1)]
!{\xbendl-}
[l(1.5)] [d(0.25)]
!{ \xbendl[0.5]}
[u(1)] [l(0.5)]
!{\xbendr[0.5]=<}
[u(1.75)] [l(2)]
!{\xunderv}
[u(1.5)] [l(1)]
!{\color{black} \xbendr-}
[u(1)] [l(1)]
!{\hcap[-2]=>}
[d(1)]
!{\xbendl \color{black}}}} \\ [-1.5cm]
\multicolumn{3}{l}{\textcolor{red} {\hspace{32pt} \xygraph{
!{0;/r1.3pc/:}
[u(2.25)] [r(3.25)]
!{\xcapv[1] =<}
[l(2.75)] [u(.75)]
!{\xbendr[-1] =<}
[d(1)] [r(1.25)]
!{\xcaph[-1] =<}
}}} \\ [-0.7cm]
		 \multicolumn{3}{r}{\hspace{145pt}$(-3,2,1)$}
		 
	\end{tabular}
	&
	\begin{tabular}{c c c}
		\multicolumn{3}{l}{$c$-preons, $D^{1/2}_{-\frac{1}{2} \frac{1}{2}}$} \\ [1.58cm]
		  & &\hspace{-63pt} $c$ \\ [-1.3cm]
		 \multicolumn{3}{l}{\xygraph{
!{0;/r1.3pc/:}
!{\xunderv=>}
[u(0.5)] [l(1)]
!{\xbendl}
[u(2)]
!{\hcap[-2]}
!{\xbendr-}
[r(1)]
!{\xbendr}
[u(2)]
!{\hcap[2]}
[l(1)]
!{\xbendl-}}} \\ [-1.2cm]
\multicolumn{3}{l}{\textcolor{red}{\hspace{39pt} \xygraph{
!{0;/r1.3pc/:}
[d(0.75)] [l(0.75)]
!{\xcaph[-1]=<}
}}} \\ [-0.7cm]
		 \multicolumn{3}{r}{\hspace{145pt}$(-1,0,1)$}
		 
	\end{tabular} \\ [-0.1cm] \hline
\begin{tabular}{c c c}
		\multicolumn{3}{l}{$d$-quarks, $D^{3/2}_{\frac{3}{2} -\frac{1}{2}} \sim ab^2$} \\ [0.2cm]
		 & $\hspace{19pt} b$ & \\ [0.4cm]
		 $\hspace{28pt} a_i$ & & $\hspace{-56pt} b$ \\ [-2.8cm]
		 \multicolumn{3}{l}{\xygraph{
!{0;/r1.3pc/:}
!{\xoverh }
[u(1)] [l(0.5)]
!{\color{black} \xbendu}
[l(2)]
!{\vcap[2]=<}
!{\xbendd- \color{black}}
[d(1)] [r(0.5)]
!{ \xoverv}
[u(0.5)] [r(1)]
!{ \xbendr }
[u(2)]
!{\hcap[2]=<}
[l(1)]
!{\xbendl- }
[u(0.5)] [l(3)]
!{\xoverv}
[u(1.5)] [l(1)]
!{\color{black} \xbendr- }
[l(1)] [u(1)]
!{\hcap[-2]=<}
[d(1)]
!{\xbendl  \color{black}}
[r(2)] [u(0.5)]
!{\xcaph- =>}}} \\[-1.6cm]
\multicolumn{3}{l}{\hspace{42pt} \textcolor{red}{\xygraph{
!{0;/r1.3pc/:}
[u(1.75)] [l(1.5)]
!{\xbendd[-1]=<}
[u(0.75)] [r(1)]
!{\xbendl[-1]=<}
[d(1)] [l(2.25)]
!{\xcaph[-1]=>}
}}} \\ [-0.7cm]
		 \multicolumn{3}{r}{\hspace{140pt}$(3,-2,1)$}
		 
	\end{tabular}
	&
	\begin{tabular}{c c c}
		\multicolumn{3}{l}{$b$-preons, $D^{1/2}_{\frac{1}{2} -\frac{1}{2}}$} \\ [1.58cm]
		  & &\hspace{-63pt} $b$ \\ [-1.3cm]
		 \multicolumn{3}{l}{\xygraph{
!{0;/r1.3pc/:}
!{\xoverv=<}
[u(0.5)] [l(1)]
!{\xbendl}
[u(2)]
!{\hcap[-2]}
!{\xbendr-}
[r(1)]
!{\xbendr}
[u(2)]
!{\hcap[2]}
[l(1)]
!{\xbendl-}}} \\ [-1.5cm]
\multicolumn{3}{l}{\hspace{40pt} \textcolor{red}{\xygraph{
!{0;/r1.3pc/:}
[u(0.75)] [l(0.75)]
!{\xcaph[1]=>}
}}} \\ [-0.1cm]
		 \multicolumn{3}{r}{\hspace{145pt}$(1,0,-1)$}
		 
	\end{tabular} \\ [-0.1cm] \hline
\begin{tabular}{c c c}
		\multicolumn{3}{l}{$u$-quarks, $D^{3/2}_{-\frac{3}{2} -\frac{1}{2}} \sim cd^2$} \\ [0.2cm]
		 & $\hspace{19pt} d$ & \\ [0.4cm]
		 $\hspace{28pt} c_i$ & & $\hspace{-56pt} d$ \\ [-2.8cm]
		 \multicolumn{3}{l}{\xygraph{
!{0;/r1.3pc/:}
!{\xunderh }
[u(1)] [l(0.5)]
!{\color{black} \xbendu}
[l(2)]
!{\vcap[2] =<}
!{\xbendd- \color{black}}
[d(1)] [r(0.5)]
!{\xunderv}
[u(0.5)] [r(1)]
!{\xbendr}
[u(2)]
!{\hcap[2] =<}
[l(1)]
!{\xbendl-}
[u(0.5)] [l(3)]
!{\xunderv}
[u(1.5)] [l(1)]
!{\color{black} \xbendr-}
[l(1)] [u(1)]
!{\hcap[-2]=<}
[d(1)]
!{\xbendl \color{black}}
[r(2)] [u(0.5)]
!{\xcaph- =>}}} \\[-1.75cm]
\multicolumn{3}{l}{\hspace{44pt}\textcolor{red}{\xygraph{
!{0;/r1.3pc/:}
[u(1.75)] [l(1.5)]
!{\xbendd[-1]=>}
[u(0.75)] [r(1)]
!{\xbendl[-1]=>}
[d(1)] [l(2.25)]
!{\xcaph[-1]=<}
}}} \\ [-0.7cm]
		 \multicolumn{3}{r}{\hspace{140pt}$(-3,-2,1)$}
		 
	\end{tabular}
	&
	\begin{tabular}{c c c}
		\multicolumn{3}{l}{$d$-preons, $D^{1/2}_{-\frac{1}{2} -\frac{1}{2}}$} \\ [1.58cm]
		  & &\hspace{-63pt} $d$ \\ [-1.3cm]
		 \multicolumn{3}{l}{\xygraph{
!{0;/r1.3pc/:}
!{\xunderv=<}
[u(0.5)] [l(1)]
!{\xbendl}
[u(2)]
!{\hcap[-2]}
!{\xbendr-}
[r(1)]
!{\xbendr}
[u(2)]
!{\hcap[2]}
[l(1)]
!{\xbendl-}}} \\ [-1.5cm]
\multicolumn{3}{l}{\hspace{39pt} \textcolor{red}{\xygraph{
!{0;/r1.3pc/:}
[u(0.75)] [l(0.75)]
!{\xcaph[1]=<}
}}} \\ [-0.1cm]
		 \multicolumn{3}{r}{\hspace{140pt}$(-1,0,-1)$} \\		 
	\end{tabular} \\ [-0.43cm]
	\multicolumn{2}{c}{The clockwise and counterclockwise arrows are given opposite weights $(\mp 1)$ respectively.} \\ [-0.51cm]
	\multicolumn{2}{c}{The \scriptsize $\binom{\text{rotation}}{\text{writhe}}$ \normalsize charge is measured by the sum of the weighted \scriptsize $\binom{\text{black}}{\text{red}}$ \normalsize arrows.} \\
\end{tabular}
\end{center}
\end{figure}


\section{Presentation of the Model in the Preon Representation$^{(3)}$}

The knot representation of $D^j_{mm'}$ by (2.1) as a function of $(a,b,c,d)$ and $(n_a,n_b,n_c,n_d)$ implies the constraints (7.1), (7.2), (7.3) on the exponents in the following way:
\begin{IEEEeqnarray}{rCl}
n_a+n_b+n_c+n_d & = & 2j \label{n2j}\\ 
n_a+n_b - n_c - n_d & =  & 2m \label{n2m}\\
n_a-n_b+n_c-n_d & = & 2m' \label{n2m'}
\end{IEEEeqnarray}
The two relations defining the quantum kinematics and giving physical meaning to $D^j_{mm'}$, namely (1.2) and (1.4):
\begin{equation*}
(j,m,m') = \frac{1}{2}(N,w, r+o) \quad \text{field (flux loop) description} \tag{1.2}
\end{equation*}
and
\begin{equation*}
(j,m,m') = 3(t, -t_3, -t_0)_L \quad \text{particle description} \tag{1.4}
\end{equation*}
imply two complementary interpretations of the relations \eqref{n2j}--\eqref{n2m'}. By (1.2) one has a field description $(N, w, \tilde{r})$ of the quantum state $(j, m, m')$ as follows
\begin{equation}
\left.
{\arraycolsep=1.2pt
\begin{array}{rl}
N &= n_a + n_b + n_c + n_d \\
w &= n_a + n_b - n_c - n_d  \\
\tilde{r} \equiv r+o &= n_a - n_b + n_c - n_d
\end{array}
}
\quad \right\} \text{field (flux loop) description} \label{field (flux loop) description}
\end{equation}
In the last line of \eqref{field (flux loop) description}, where $\tilde{r} \equiv r+o$ and $o$ is the parity index, $\tilde{r}$ has been termed ``the quantum rotation," and $o$ the ``zero-point rotation." 
\newline

By (1.4) one has a particle description $(t, t_3, t_0)$ of the same quantum state $(j, m, m')$.
\begin{equation}
\left.
{\arraycolsep=1.2pt
\begin{array}{rl}
t &= \frac{1}{6} (n_a + n_b +n_c +n_d)  \\
t_3 &= -\frac{1}{6}(n_a + n_b - n_c - n_d)  \\
t_0 &= -\frac{1}{6}(n_a - n_b + n_c - n_d) 
\end{array}
}
\quad \right\} \text{particle description} \label{particle_description}
\end{equation}
\emph{In \eqref{particle_description}, $(t, t_3, t_0)$ are to be read as SLq(2) indices agreeing with standard notation only at $j = \frac{3}{2}$. Then in general $t_3$ measures writhe charge, $t_0$ measures rotation hypercharge and $t$ measures the total preon population or the total number of crossings.}
\newline

\section{Particle--Field Complementarity Restated}
In the flux loop description equations \eqref{field (flux loop) description}, the numerical coefficients may be replaced by $(N_p, w_p, \tilde{r}_p)$ describing the preons as follows: \vspace{-0.5em} 
\begin{center}
\vspace{-0.5em}
\renewcommand{\tabcolsep}{3pt}
\begin{tabular}{l r}
\begin{tabular}{l l r l}
 & $N = \sum_p n_p N_p$ & & (8.1) \\ 
 & $ w = \sum_p n_p w_p $\hspace{-15pt} \hspace{-5pt} & & (8.2) \hspace{10pt} where  \\
 & $\tilde{r} = \sum_p n_p \tilde{r}_p$ & & (8.3) \hspace{50pt} \\
\end{tabular}
&
\begin{tabular}{r | r r r}
\hline \hline
$p$ & $N_p$ & $w_p$ & $\tilde{r}_p$ \\
\hline
$a$ & 1 & 1 & 1 \\
$b$ & 1 & 1 & $-1$ \\
$c$ & 1 & $-1$ & 1 \\
$d$ & 1 & $-1$ & $-1$ \\
\hline
\end{tabular}
\end{tabular}
\vspace{0.5em}
\end{center} \vspace{-0.3em}
and in the particle description equations (7.5), the numerical coefficients may be replaced by $(t_p, t_{3_p}, t_{0_p})$ describing the preons as follows:
\begin{center}
\vspace{-0.5em}
\renewcommand{\tabcolsep}{3pt}
\begin{tabular}{l r}
\begin{tabular}{l l r l}
$t = \sum_p n_p t_p$ & & & (8.4) \hspace{50pt} \\
$ t_3 = \sum_p n_p t_{3_p} $  &\hspace{-15pt} \hspace{-5pt} & & (8.5) \hspace{10pt} where \\
$ t_0 = \sum_p n_p t_{0_p} $  & & & (8.6) \\
\end{tabular}
& \normalsize
\begin{tabular}{r | r r r }
\hline \hline
$p$ & $t_p$ & $t_{3_p}$ & $t_{0_p}$ \\
\hline
$a$ & $\frac{1}{6}$ & $-\frac{1}{6}$ & $-\frac{1}{6}$ \\
$b$ & $\frac{1}{6}$ & $-\frac{1}{6}$ & $\frac{1}{6}$ \\
$c$ & $\frac{1}{6}$ & $\frac{1}{6}$ & $-\frac{1}{6}$  \\
$d$ & $\frac{1}{6}$ & $\frac{1}{6}$ & $\frac{1}{6}$  \\
\hline
\end{tabular}
\end{tabular}
\vspace{0.5em}
\end{center} \vspace{-0.3em}
\addtocounter{equation}{6}
Since $r=0$ for preonic loops, $o$ plays the role of a quantum rotation for preons:
\begin{align}
\tilde{r}_p = r_p + o_p = o_p \qquad \qquad p = (a, b, c, d)
\end{align}
For the elementary fermions presently observed,
\begin{align}
\tilde{r} = r + 1. 
\end{align}
\emph{The formal algebraic relations (8.1)--(8.6) express properties of the higher representations of the SLq(2) algebra as additive compositions of the fundamental representation.} \emph{The quantum state is now defined by $(n_a, n_b, n_c, n_d)$, the preon populations. It is as well still defined by $(j, m, m')$, the SLq(2) representation, and by the complementary knot $(N, w, \tilde{r})$ and particle descriptions, $(t, t_3, t_0)$}. All of these descriptions impose the same quantum kinematics.

\section{Interpretation of the Complementary Equations Continued}
We now present a particle interpretation of equations (7.4)
\begin{align*}
N = n_a + n_b + n_c + n_d \tag{7.4$N$} \\
w = n_a + n_b - n_c - n_d \tag{7.4$w$} \\
\tilde{r} = n_a - n_b + n_c - n_d \tag{7.4$\tilde{r}$}
\end{align*}

\emph{Equation (7.4$N$) states that the number of crossings, $N$, equals the total number of preons, $N'$. Since we assume that the preons are fermions, the knot describes a fermion or a boson depending on whether the number of crossings is odd or even.} Viewed as a knot, a fermion becomes a boson when the number of crossings is changed by attaching or removing a curl. This picture is consistent with the view of a curl as an opened preon loop, viewed as a twisted loop. Each counterclockwise or clockwise classical curl corresponds to a preon creation operator or antipreon creation operator respectively.
\newline

Since $a$ and $d$ are creation operators for antiparticles with opposite charge and hypercharge, while $b$ and $c$ are neutral antiparticles with opposite values of the hypercharge, we may introduce the preon numbers
\begin{align}
\nu_a &= n_a -n_d \\
\nu_b &= n_b - n_c
\end{align}
Then (7.4$w$) and (7.4$\tilde{r}$) may be rewritten in terms of preon numbers as
\begin{align}
&\nu_a + \nu_b = w \hspace{3pt} (= -6t_3) \\
&\nu_a - \nu_b = \tilde{r} \hspace{3pt} (=-6t_0)
\end{align}
By (9.3) and (9.4) the conservation of the preon numbers and of the charge and hypercharge is equivalent to the conservation of the writhe and rotation, which are topologically conserved at the 2d-classical level. In this respect, these quantum conservation laws for preon numbers correspond to the classical conservation laws for writhe and rotation.
\newline

\section{Measure of Charge by SU(2) $\times$ U(1) and by SLq(2)} 
The SU(2)$\times$U(1) measure of charge requires the assumption of fractional charges for the quarks. The SLq(2) measure requires the replacement of the fundamental charge $(e)$ for charged leptons by a new fundamental charge $(e/3)$ for charged preons but does not require fractional charges for quarks.$^{(5)}$
\newline

The SLq(2), or $(j,m,m')$ measure, has a direct physical interpretation since $2j$ is the total number of preonic sources, while $2m$ and $2m'$ respectively measure the numbers of writhe and rotation sources of preonic charge.$^{(3)}$ Since $N$, $w$, and $r$ all measure the handedness of the source, charge is also measured by the handedness of the source.
\newline

If neutral unknotted flux tubes predated the particles, and the particles were initially formed by the knotting of the neutral flux tubes of energy-momentum, then the simplest fermions that could have formed must have had 3 crossings and therefore three preons. \emph{The electric charge of the resultant trefoil or of any composite of preons would then be a measure of the handedness generated by the knotting of an original unknotted flux loop.}
\newline

The total SLq(2) charge sums the signed clockwise and counterclockwise turns that any energy-momentum current makes both at the crossings and in making one circuit of the 2d-projected knot. This measure of charge, ``knot charge", which is suggested by the leptons and quarks, appears more fundamental than the electroweak isotopic measure that originated in the neutron-proton system, since it reduces the concept of charge to the handedness of the source and in this way reduces charge to a topological concept similar to the way energy-momentum is geometrisized by the curvature of spacetime. \newline

If charge is a measure of handedness in the energy-momentum current, then there is also a natural correspondence between (writhe, rotation) charge and (spin, orbital) angular momentum.

\section{Physical Interpretation of Corresponding Quantum States}

The point particle $(N', \nu_a, \nu_b)$ representation and the flux loop $(N, w, \tilde{r})$ complementary representation are related by
\begin{align}
\tilde{D}^{N'}_{\nu_a, \nu_b} = \sum_{N,w,r} \delta(N',N)\delta(\nu_a+\nu_b, w)\delta(\nu_a - \nu_b, \tilde{r}) D^{N/2}_{\frac{w}{2} \frac{\tilde{r}}{2}}
\end{align}
Since one may interpret the elements $(a,b,c,d)$ of the SLq(2) algebra as creation operators for either preonic particles or flux loops, the $D^j_{mp}$ may be interpreted as a creation operator for a composite particle composed of either preonic particles $(N', \nu_a, \nu_b)$ or flux loops $(N, w, \tilde{r})$ . \emph{These two complementary views of the same particle may be reconciled as describing $N$-preon systems bound by a knotted field having $N$-crossings with the preons at the crossings as illustrated in Figure 2 for $N=3$.} In the limit where the three outside lobes become small or infinitesimal compared to the central circuit, the resultant structure will resemble a three particle system tied together by a string. 
\begin{center}
\normalsize
\begin{tabular}{c | c}
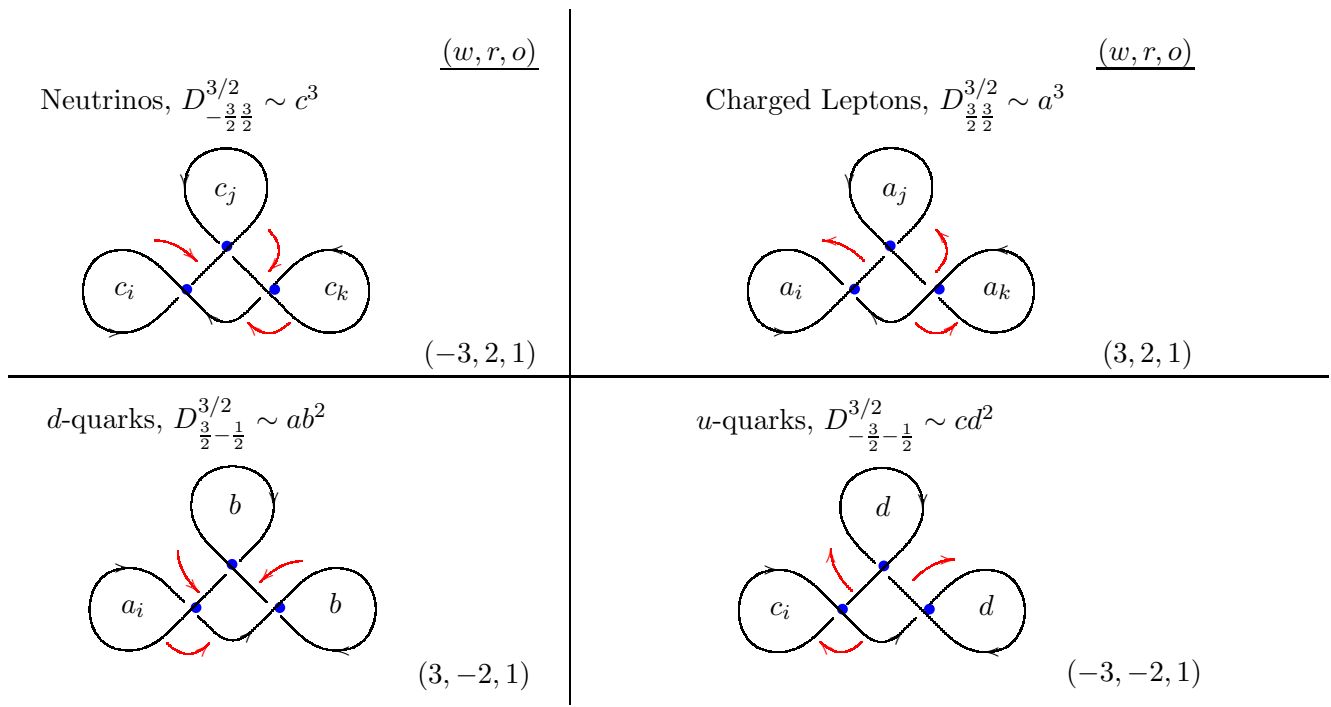
	\multicolumn{2}{c}{\textbf{Figure 2:} Leptons and Quarks Pictured as Three Preons Bound by a Trefoil Field} \\ [-0.0cm]
	\begin{tabular}{c c c}
		\multicolumn{3}{r}{ \ul{$(w, r, o)$}} \\ [-0.3cm]
		\multicolumn{3}{l}{Neutrinos, $D^{3/2}_{-\frac{3}{2} \frac{3}{2}} \sim c^3$} \\ [0.2cm]
		 & $\hspace{17pt} c_j$ & \\ [-.22 cm]
		 \multicolumn{3}{c}{\hspace{-1.655 cm}\vspace{0.22 cm}\textcolor{blue} {\scalebox{2}{\Huge .}}} \\ [-0.6cm]
		 $\hspace{28pt} c_i$ {\hspace{.33cm}\textcolor{blue} {\scalebox{2}{\Huge .}}} \hspace{-1cm} & & {\hspace{-2.65cm}\textcolor{blue} {\scalebox{2}{\Huge .}}} \hspace{2.15cm} $\hspace{-56pt} c_k$ \\ [-2.8cm]
		 \multicolumn{3}{l}{\xygraph{
!{0;/r1.3pc/:}
!{\xunderh }
[u(1)] [l(0.5)]
!{\xbendu}
[l(2)]
!{\vcap[2]=>}
!{\xbendd-}
[d(1)] [r(0.5)]
!{\xoverh}
[d(0.5)]
!{\xbendr}
[u(2)]
!{\hcap[2]=>}
[l(1)]
!{\xbendl-}
[l(1.5)] [d(0.25)]
!{\xbendl[0.5]}
[u(1)] [l(0.5)]
!{\xbendr[0.5]=<}
[u(1.75)] [l(2)]
!{\xunderv}
[u(1.5)] [l(1)]
!{\xbendr-}
[u(1)] [l(1)]
!{\hcap[-2]=>}
[d(1)]
!{\xbendl}}} \\[-1.5cm]
\multicolumn{3}{l}{\textcolor{red} {\hspace{32pt} \xygraph{
!{0;/r1.3pc/:}
[u(2.25)] [r(3.25)]
!{\xcapv[1] =<}
[l(2.75)] [u(.75)]
!{\xbendr[-1] =<}
[d(1)] [r(1.25)]
!{\xcaph[-1] =<}
}}} \\ [-0.7cm]
		 \multicolumn{3}{r}{\hspace{145pt}$(-3,2,1)$}
		 
	\end{tabular}
&
\begin{tabular}{c c c}
		\multicolumn{3}{r}{ \ul{$(w, r, o)$}} \\ [-0.3cm]
		\multicolumn{3}{l}{Charged Leptons, $D^{3/2}_{\frac{3}{2} \frac{3}{2}} \sim a^3$} \\ [0.2cm]
		 & $\hspace{17pt} a_j$ & \\ [-.22 cm]
		 \multicolumn{3}{c}{\hspace{-1.58 cm}\vspace{0.22 cm}\textcolor{blue} {\scalebox{2}{\Huge .}}} \\ [-0.6cm]
		 $\hspace{28pt} a_i$ {\hspace{.33cm}\textcolor{blue} {\scalebox{2}{\Huge .}}} \hspace{-1cm}  & & {\hspace{-2.65cm}\textcolor{blue} {\scalebox{2}{\Huge .}}} \hspace{2.15cm} $\hspace{-58pt} a_k$ \\ [-2.8cm]
		 \multicolumn{3}{l}{\xygraph{
!{0;/r1.3pc/:}
!{\xoverh }
[u(1)] [l(0.5)]
!{\xbendu}
[l(2)]
!{\vcap[2]=>}
!{\xbendd-}
[d(1)] [r(0.5)]
!{\xunderh}
[d(0.5)]
!{\xbendr }
[u(2)]
!{\hcap[2]=>}
[l(1)]
!{\xbendl-}
[l(1.5)] [d(0.25)]
!{\xbendl[0.5]}
[u(1)] [l(0.5)]
!{\xbendr[0.5]=<}
[u(1.75)] [l(2)]
!{\xoverv}
[u(1.5)] [l(1)]
!{\xbendr-}
[u(1)] [l(1)]
!{\hcap[-2]=>}
[d(1)]
!{\xbendl}}} \\ [-1.6cm]
\multicolumn{3}{l}{\hspace{33pt} \textcolor{red}{
\xygraph{
!{0;/r1.3pc/:}
[u(2.25)] [r(3.25)]
!{\xcapv[1] =>}
[l(2.75)] [u(.75)]
!{\xbendr[-1] =>}
[d(1)] [r(1.25)]
!{\xcaph[-1] =>}
}}} \\ [-0.7cm]
		 \multicolumn{3}{r}{\hspace{150pt}$(3,2,1)$}
		 
	\end{tabular} \\ [-0.1cm] \hline
	\begin{tabular}{c c c}
		\multicolumn{3}{l}{$d$-quarks, $D^{3/2}_{\frac{3}{2} -\frac{1}{2}} \sim ab^2$} \\ [0.2cm]
		 & $\hspace{19pt} b$ & \\ [-.22 cm]
		  \multicolumn{3}{c}{\hspace{-1.52 cm}\vspace{0.22 cm}\textcolor{blue} {\scalebox{2}{\Huge .}}} \\ [-0.6cm]
		 $\hspace{28pt} a_i$ {\hspace{.33cm}\textcolor{blue} {\scalebox{2}{\Huge .}}} \hspace{-1cm}  & & {\hspace{-2.65cm}\textcolor{blue} {\scalebox{2}{\Huge .}}} \hspace{2.15cm} $\hspace{-56pt} b$ \\ [-2.8cm]
		 \multicolumn{3}{l}{\xygraph{
!{0;/r1.3pc/:}
!{\xoverh }
[u(1)] [l(0.5)]
!{\xbendu}
[l(2)]
!{\vcap[2]=<}
!{\xbendd-}
[d(1)] [r(0.5)]
!{\xoverv}
[u(0.5)] [r(1)]
!{\xbendr}
[u(2)]
!{\hcap[2] =<}
[l(1)]
!{\xbendl-}
[u(0.5)] [l(3)]
!{\xoverv}
[u(1.5)] [l(1)]
!{\xbendr-}
[l(1)] [u(1)]
!{\hcap[-2]=<}
[d(1)]
!{\xbendl}
[r(2)] [u(0.5)]
!{\xcaph- =>}}} \\ [-1.6cm]
\multicolumn{3}{l}{\hspace{42pt} \textcolor{red}{\xygraph{
!{0;/r1.3pc/:}
[u(1.75)] [l(1.5)]
!{\xbendd[-1]=<}
[u(0.75)] [r(1)]
!{\xbendl[-1]=<}
[d(1)] [l(2.25)]
!{\xcaph[-1]=>}
}}} \\ [-0.7cm]
		 \multicolumn{3}{r}{\hspace{140pt}$(3,-2,1)$}
		 
	\end{tabular}
	&
\begin{tabular}{c c c}
		\multicolumn{3}{l}{$u$-quarks, $D^{3/2}_{-\frac{3}{2} -\frac{1}{2}} \sim cd^2$} \\ [0.2cm]
		 & $\hspace{19pt} d$ & \\ [-.22 cm]
		  \multicolumn{3}{c}{\hspace{-1.75 cm}\vspace{0.22 cm}\textcolor{blue} {\scalebox{2}{\Huge .}}} \\ [-0.6cm]
		 $\hspace{28pt} c_i$ {\hspace{.33cm}\textcolor{blue} {\scalebox{2}{\Huge .}}} \hspace{-1cm}  & & {\hspace{-2.9cm}\textcolor{blue} {\scalebox{2}{\Huge .}}} \hspace{2.15cm} $\hspace{-56pt} d$ \\ [-2.8cm]
		 \multicolumn{3}{l}{\xygraph{
!{0;/r1.3pc/:}
!{\xunderh }
[u(1)] [l(0.5)]
!{\xbendu}
[l(2)]
!{\vcap[2]=<}
!{\xbendd-}
[d(1)] [r(0.5)]
!{\xunderv}
[u(0.5)] [r(1)]
!{\xbendr}
[u(2)]
!{\hcap[2]=<}
[l(1)]
!{\xbendl-}
[u(0.5)] [l(3)]
!{\xunderv}
[u(1.5)] [l(1)]
!{\xbendr-}
[l(1)] [u(1)]
!{\hcap[-2]=<}
[d(1)]
!{\xbendl}
[r(2)] [u(0.5)]
!{\xcaph- =>}}} \\ [-1.75cm]
\multicolumn{3}{l}{\hspace{44pt}\textcolor{red}{\xygraph{
!{0;/r1.3pc/:}
[u(1.75)] [l(1.5)]
!{\xbendd[-1]=>}
[u(0.75)] [r(1)]
!{\xbendl[-1]=>}
[d(1)] [l(2.25)]
!{\xcaph[-1]=<}
}}} \\ [-0.7cm]
		 \multicolumn{3}{r}{\hspace{140pt}$(-3,-2,1)$}
		 
	\end{tabular} \\ [-0.2cm]
\multicolumn{2}{c}{The preons conjectured to be present at the crossings are suggested by the blue dots at the crossings.}
\end{tabular}
\end{center}
\emph{The physical models suggested by Fig. 2 may be further studied in the context of gravitational and gluon binding with the aid of the preon Lagrangian given in reference 3}. The required superstrong binding force might also be provided by an extension of the Schwinger conjecture, not that the quarks are dyons, but by a similar conjecture about the preons. There is no experimental guidance at these conjectured energies.
\newline

\section{Alternate Interpretation}
In the model suggested by Fig. 2 the parameters of the preons and the parameters of the flux loops are to be understood as codetermined. On the other hand, in an alternative interpretation of complementarity, the hypothetical preons conjectured to be present in Figure 2 carry no independent degrees of freedom and may simply describe \emph{concentrations of energy and momentum at the crossings of the flux tube}. In this interpretation of complementarity, $(t, t_3, t_0)$ and $(N,w, \tilde{r})$ are just two ways of describing the same quantum trefoil of field. In this picture the preons are bound, i.e. they do not appear as free particles. This view of the elementary particles as non-singular lumps of field has also been described as a unitary field theory.
\newline

\section{Lower Representations}
We have so far considered the states $j = 3, \frac{3}{2}, \frac{1}{2}$ representing electroweak vectors, leptons and quarks, and preons, respectively. We finally consider the states $j=1$ and $j=0$. Here we shall not examine the higher $j$ states.
\newline

In the adjoint representation $j=1$, the particles are the vector bosons by which the preons interact and there are two crossings. These vectors are different from the $j=3$ vectors by which the leptons and quarks interact.
\newline

If $j=0$, the indices of the quantum knot are
\begin{align}
(j,m,m') = (0,0,0)
\end{align}
and by the rule (1.2) for interpreting the knot indices on the left chiral fields
\begin{align}
\frac{1}{2}(N,w,\tilde{r}) = (j, m, m') &= (0, 0, 0) 
\end{align}
Then the $j=0$ quantum states correspond to classical loops with no crossings $(N=0)$ just as preon states correspond to classical twisted loops with one crossing. Since $N=0$, the $j=0$ states also have no preonic sources of charge and therefore no electroweak interaction.  \emph{It is possible that these }$j=0$ \emph{hypothetical quantum states are realized as (electroweak non-interacting) loops of field flux with} $w=0$, $\tilde{r}= r+o = 0$\emph{, and }$r = \pm1$, $o =\mp1$ \emph{ i.e. with the topological rotation }$r=\pm1$. The two states $(r, o) = (+1, -1)$ and $(-1, +1)$ are to be understood as quantum mechanically coupled.
\newline

If, as we are assuming, the leptons and quarks with $j= \frac{3}{2}$ correspond to 2d projections of knots with three crossings, and if the heavier preons with $j= \frac{1}{2}$ correspond to 2d projections of twisted loops with one crossing, then if the $j=0$ states correspond to 2d projections of simple loops, one might ask if these particles with no electroweak interactions are smaller and heavier than the preons, and are among the candidates for ``dark matter." If these $j=0$ particles predated the $j=\frac{1}{2}$ preons, one may refer to them as ``yons" as suggested by the term ``ylem" for primordial matter.
\newline

\section{Speculations about an earlier universe and dark matter} \vspace{-0.8em}
One may speculate about an earlier universe before leptons and quarks had appeared, when there was no charge, and when energy and momentum existed only in the SLq(2) $j=0$ neutral state as simple loop currents of gravitational energy-momentum. Then the gravitational attraction would bring some pairs of opposing loops close enough to permit the transition from two $j=0$ loops into two opposing $j=\frac{1}{2}$ twisted loops. A possible geometric scenario for the transformation of two simple loops of current (yons) with opposite rotations into two $j=\frac{1}{2}$ twisted loops of current (preons) is shown in Fig. 3. To implement this scenario one would expect to go beyond the electroweak dynamics. Without attempting to do this, one notes according to Fig. 3 that the fusion of two yons may result in a doublet of preons as twisted loops.
\newline

\begin{center}
\normalsize
\begin{tabular}{c  c  c}
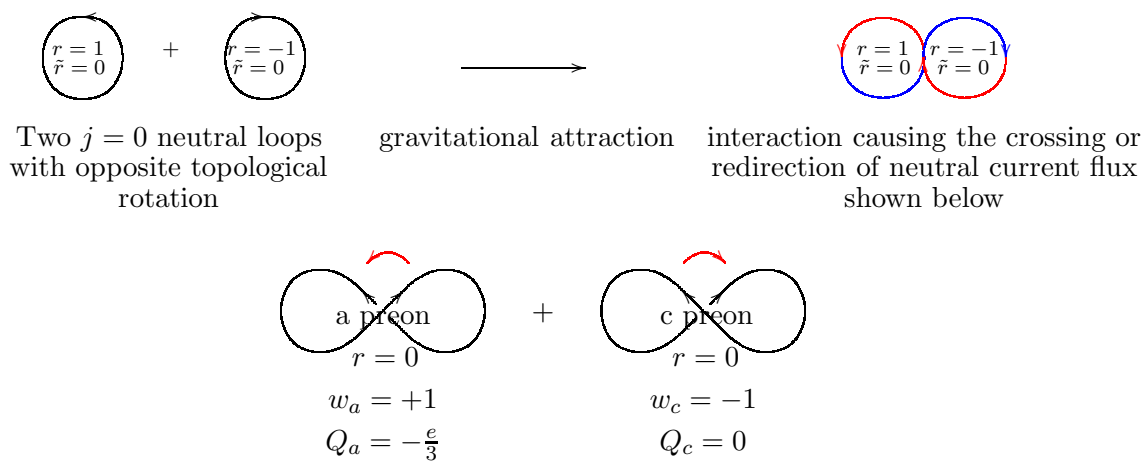
\multicolumn{3}{c}{ \textbf{Figure 3:} Creation of Preons as Twisted Loops} \\
\vspace{-3pt}
\xygraph{
!{0;/r1.3pc/:}
!{\hcap[2]=>}
!{\hcap[-2]}} \hspace{3pt} 
\xygraph{
!{0;/r1.3pc/:}
!{\hcap[2]}
!{ \hcap[-2]=< }} &
\xygraph{
!{0;/r1.3pc/:}
[d(1.25)]
!{\xcaph[3]=<@(0)}} & \hspace{-4pt}
\xygraph{
!{0;/r1.3pc/:}
[d(1)]
!{\color{red} \vcap[2]=> \color{blue}}
!{\vcap[-2]}} \hspace{-7pt} 
\xygraph{
!{0;/r1.3pc/:}
[d(1)]
!{\color{blue} \vcap[2]=<}
!{\color{red} \vcap[-2]=< \color{black}}}\\ [-1.7cm]
\hspace{4pt}\scriptsize{$r=1 \hspace{20pt}+ \hspace{16pt}r=-1$} & & \scriptsize{$\hspace{3pt}r=1 \hspace{8pt} r=-1$} \\ [-0.7cm]
\hspace{0pt}\scriptsize{$\tilde{r}=0 \hspace{32pt} \hspace{16pt}\tilde{r}=0$} & & \scriptsize{$\hspace{0pt}\tilde{r}=0 \hspace{10pt} \tilde{r}=0$} \\
Two $j=0$ neutral loops & gravitational attraction & interaction causing the crossing or  \\ [-0.55cm]
 with opposite topological & & redirection of neutral current flux \\ [-0.55cm]
  rotation & & shown below \\ [-0.55cm]
\end{tabular}
\end{center}
\begin{center}
\begin{tabular}{c c}
\textcolor{red}{\xygraph{
!{0;/r1.3pc/:}
[u(0.75)] [l(0.75)]
!{\xcaph[1]=>}
}} &
\textcolor{red}{\xygraph{
!{0;/r1.3pc/:}
[u(0.75)] [l(0.75)]
!{\xcaph[1]=<}
}} \\ [-1cm]
\xygraph{
!{0;/r1.3pc/:}
!{\xoverv=<}
[u(0.5)] [l(1)]
!{\xbendl}
[u(2)]
!{\hcap[-2]}
!{\xbendr-}
[r(1)]
!{\xbendr}
[u(2)]
!{\hcap[2]}
[l(1)]
!{\xbendl-}} &
\xygraph{
!{0;/r1.3pc/:}
!{\xunderv=<}
[u(0.5)] [l(1)]
!{\xbendl}
[u(2)]
!{\hcap[-2]}
!{\xbendr-}
[r(1)]
!{\xbendr}
[u(2)]
!{\hcap[2]}
[l(1)]
!{\xbendl-}} \\ [-1.1cm]
\multicolumn{2}{c}{+} \\ [-0.95cm]
\hspace{-95pt} & \hspace{-95pt} \\ [-0.9cm]
a preon & \hspace{-1pt} c preon \\ [-0.4cm]
\hspace{2pt}$r=0$ & \hspace{2pt}$r=0$ \\ [-0.4cm]
$w_a = +1$ & $w_c=-1$ \\ [-0.4cm]
$ Q_a = -\frac{e}{3}$ & $ Q_c = 0$  \\
\end{tabular}
\end{center}
In the scenario suggested by Figure 3 the opposing states are quantum mechanically entangled and may undergo gravitational exchange scattering. 
%
\newline

The $\binom{c}{a}$ doublet of Fig. 3 is similar to the Higgs doublet which is independently required to be a SLq(2) singlet $(j=0)$ and a SU(2) charge doublet $(t=\frac{1}{2})$ by the mass term of the Lagrangian described in reference 3. Since the Higgs mass is also the inertial mass, one expects a fundamental connection with the gravitational field at this point.
\newline \vspace{-0.6em}

A fraction of the preons $(j= \frac{1}{2})$ produced by the fusion of two yons would in turn combine to form two preon $(j=1)$ and then three preon $(j = \frac{3}{2})$ states with two and three crossings respectively. The $j = \frac{3}{2}$ states would be recognized in the present universe as leptons and quarks. Since, however, the $j = \frac{1}{2}$ and $j=1$ particles with one and two crossings, respectively, are not topologically stable in three dimensions and can relapse into a $j=0$ state with no crossings, the building up process from yons does not produce topologically stable particles before the $j=\frac{3}{2}$ leptons and quarks with three crossings are reached.
\newline \vspace{-0.6em}
 
If at an early cosmological time, only a fraction of the initial gas of quantum loops was converted to preons and these in turn led to a still smaller number of leptons and quarks, then most of the mass and energy of the universe would at the present time still reside in the dark loops while charge and current and visible mass would be confined to structures composed of leptons and quarks. \emph{In making experimental tests for particles of dark matter one would expect the SLq(2) $j=0$ dark loops to be different in mass from the dark neutrino trefoils where $j= \frac{3}{2}$.}
\newline

\section{A Possible Interpretation of the Parameter $q$ in the SLq(2) Model$^{(3)}$} 

In the SLq(2) extension of the standard model, as so far presented here, $q$ is regarded as a deformation parameter without a physical interpretation. Since the only physical coupling constants appearing explicitly in this model are electroweak, it is necessary to assign SU(3) indices to the preons in order to go beyond electroweak physics but the gluon couplings are then only implicit and are not completely introduced. It may be possible, however, to introduce more fully both the electroweak and the gluon couplings in the knot factors by interpreting $q$ as a physical parameter. To explore this possibility one may present SLq(2) in terms of the following 2 parameter matrix:
\begin{align}
\varepsilon_q = \begin{pmatrix}
0 & \alpha_2 \\
-\alpha_1 & 0 \\
\end{pmatrix}
\end{align}
invariant under SLq(2) as follows:
\begin{equation}
T \varepsilon_q T^t = T^t \varepsilon_q T = \varepsilon_q \label{tepsilon}
\end{equation}
where $t$ means transpose and where $T$ is a two dimensional representation of SLq(2):
\begin{align}
T = \begin{pmatrix}
a & b \\
c & d \end{pmatrix} . \label{tmatrix}
\end{align}
Then \eqref{tepsilon} with \eqref{tmatrix} generates the SLq(2) algebra (2.2) as follows:
\begin{equation}
\begin{split}
ab = qba \qquad bd = qdb \qquad ad-qbc = 1 \qquad bc &= cb \\
ac = qca \qquad cd = qdc \qquad da -q_1cb = 1 \qquad q_1 &\equiv q^{-1}
\end{split}
\end{equation}
where now
\begin{align}
q = \frac{\alpha_1}{\alpha_2} \label{alpha12}
\end{align}
If one also requires
\begin{align}
\text{det} \hspace{2pt} \varepsilon_q = 1 \label{deteps}
\end{align}
then
\begin{align}
\alpha_1 \alpha_2 = 1
\end{align}
and $\varepsilon_q$ underlies the structure of the Kauffman knot polynomial.$^{(3)}$

Using this same algebra to describe the field theory of a particle that carries two different charges, $e$ and $g$, we interpret the invariant matrix, $\varepsilon_q$, as a matrix coupling by setting
\begin{align}
(\alpha_2, \alpha_1) \hspace{4pt} \text{or} \hspace{4pt} (\alpha_1, \alpha_2) = \bigg(\frac{e}{\sqrt{\hbar c}}, \frac{g}{\sqrt{\hbar c}} \bigg ) \label{gluonchargealpha}
\end{align}
where $\alpha_1$ and $\alpha_2$ are dimensionless and $e$ and $g$ have the dimensions of an electric charge. Then by \eqref{alpha12}
\begin{align}
q = \frac{e}{g} \hspace{4pt} \text{or} \hspace{4pt} \frac{g}{e} \label{qegratio}
\end{align}
and if \eqref{deteps} is also imposed,$^{(3)}$
\begin{align}
eg = \hbar c \label{qinterpretation}
\end{align}
Here $e$, $g$, and $q$ may be running coupling constants and $e$ and $g$ may be normalized to agree with experiment at hadronic or higher energies. If $e$ increases with energy and $g$ decreases with energy according to asymptotic freedom, $q$ may become very large or very small at the high energies where the interaction and mass terms become relevant for fixing the three particle bound states representing charged leptons, neutrinos and quarks. Although there is currently no experimental data suggesting the interpretation of $q$ as a particular function of an $e$ and a $g$, such a relation could be explored since $e$, $g$, and $q$ can be independently measured.

If $g$ is interpreted not as gluon charge but as magnetic charge in \eqref{gluonchargealpha} then \eqref{qinterpretation} resembles the Dirac equation connecting electric and magnetic charge. If one also goes beyond electroweak to assume that the preons are dyons and that the binding force is magnetic, then $g$ may measure the magnetic charge of the preons and the deformation parameter $q$ may measure the deformation of the electroweak model required by the magnetic charge to pass from the standard model to the knot extended standard model. At the present time $q$ is simply an unknown that parametrizes the model.
\newline

The interpretation of
\begin{align*}
q &= \frac{e}{g} \quad \text{or} \quad \frac{g}{e} \tag{15.9} \\
\text{and} \quad eg &= \hbar c \tag{15.10}
\end{align*}
is here described in terms of theoretical concepts as expressed in the standard model and suggested by experiments carried out at hadronic energies. If one goes to much higher energies, then the theoretical concepts and interpretation of \eqref{qegratio} and \eqref{qinterpretation} may change since the gravitational field at very short distances as well as the dimensionality of spacetime may change$^{(9)}$, but in both the gluon and dyon interpretations the underlying symmetry assumed here is SLq(2).

\section*{Acknowledgments}
I thank J.~Smit, A.~C.~Cadavid, and J.~Sonnenschein for helpful discussion.

%
%
%
%
%
%
%
%

\end{document}